\documentclass[a4paper,12pt]{iopart}
\usepackage{graphicx}
\usepackage{epstopdf}

\begin{document}

\comment{Comment on `Series expansions from the corner transfer matrix 
renormalization group method: the hard-squares model'}
 \author{Iwan Jensen \\ }
\address{\small ARC Centre of Excellence for Mathematics and Statistics of Complex Systems,
Department of Mathematics and Statistics,
The University of Melbourne, VIC 3010, Australia}
\begin{abstract}
Earlier this year Chan extended the low-density series for the hard-squares partition function
$\kappa(z)$ to 92 terms. Here we analyse this extended series focusing on the behaviour at the
dominant singularity $z_d$ which lies on on the negative fugacity axis. We find that
the series has a confluent singularity of order 2 at $z_d$ with exponents 
$\theta=0.83333(2)$ and $\theta'= 1.6676(3)$.  We thus confirm that the exponent
$\theta$ has the exact value $\frac56$ as observed by Dhar. 
\end{abstract}

In \cite{Chan12} Y. Chan  extended the low-density series for the hard-squares
partition function $\kappa (z)$ to 92 terms. In a brief analysis of the associated
magnetisation series $M(z)=\frac{\rmd}{\rmd z} \ln \kappa (z)$ Chan found that
this series has a physical singularity at $z_c=3.79635(9)$ with exponent
$\alpha = 0.0020(17)$. This is in complete agreement with more accurate numerical
work  that has unequivocally established that the 
critical behaviour of the hard-squares model is in the Ising universality class
and hence $\alpha=0$. The current best estimate for $z_c$ is to our
knowledge $z_c=3.79625517391234(4)$ \cite{Guo02}.
More interestingly Chan gives a very accurate estimate for the dominant 
singularity of $\kappa(z)$ which happens to lie on the negative real axis 
at $z_d=-0.1193388818(6)$ with critical exponent $-\gamma = 0.171(14)$.
 Dhar \cite{Dhar83} has related the hard-squares model
to directed-site animals and analysed a 42-term hard-squares series for $\kappa(z)$ 
calculated by Baxter \etal \cite{Baxter80a} and found that the critical exponent at $z_d$ is
$\theta=\frac56$. Obviously, $\gamma = 1-\theta=\frac16$. The series was also
analysed by Guttmann \cite{Guttmann87} using various sequence extrapolation
methods (see \cite{GuttmannDA} for a review). The most accurate estimates was 
obtained using Brezinski's $\theta$ transform (this $\theta$ is unrelated to the
critical exponent) and Guttmann found $z_d=-0.1193388809(10)$ and 
$\theta \simeq 0.83338$ (no error estimate was given).

The results of Chan are thus are quite surprising in that $z_d$ is obtained to 
10 digit accuracy and $\gamma$ only to 2 digit accuracy. This immediately suggests
that the critical behaviour at $z_d$ is more complicated than assumed in Chan's 
analysis. The most obvious complication is that the singularity at $z_d$
contains confluent terms.  In this comment  we  analyse the series
for $\kappa(z)$ and demonstrate that this
is indeed the case and we show that the critical point $z_d$ is at least a double root.
This refined analysis then allows us to obtain  accurate estimates for the
exponents at $z_d$, namely, $\theta=0.83333(2)$ and $\theta'= 1.6676(3)$,
 which obviously confirms the observation by Dhar \cite{Dhar83} that $\theta =5/6$
 and suggest that  $\theta'$ could equal $2\theta$.

To estimate the singularities and exponents   of $\kappa(z)$ we  (as did Chan) use
the numerical method of differential approximants \cite{GuttmannDA}.  
We refer the interested reader to \cite{GuttmannDA} for details, 
and Chapter 8 of \cite{PolygonBook} for an overview of the method. 
Suffice to briefly say that  a $K$'th-order differential approximant to a function $F(z)$, 
for which one has derived a series expansion, is formed by determining the coefficients in the
polynomials $Q_i(z)$ and $P(z)$ of order $N_i$ and $L$, respective,
so that the solution $\tilde{F}(z)$  to the inhomogeneous differential equation
\begin{equation}\label{eq:diffapp}
\sum_{i=0}^K Q_{i}(z)(z\frac{\rmd}{\rmd z})^i \tilde{F}(z) = P(z)
\end{equation}
agrees with the series coefficients of $F(z)$ up to an order determined
by the number of unknown coefficients in (\ref{eq:diffapp}).
The possible singularities of the series appear as
the zeros $z_i$ of the polynomial $Q_K(z)$ and the associated critical
exponents $\lambda_i$ are obtained from the indicial equation. 
Note that not all roots of $Q_K$ are actual singularities
of the underlying series.

In table~\ref{tab:da21}  we list all the real zeros of $Q_3(z)$ and the associated 
exponents as obtained from a  homogeneous third order differential approximant 
with  polynomials of degree 21. The exponents were calculated
assuming that all the roots are distinct and hence of order 1. 
We immediately notice that if the two zeros close to $z_d$  (bold-faced in the table) 
are distinct they lie incredibly close to one another. A more likely scenario is that
the root at $z_d$ is of order at least 2.  If we assume that the singularity has order two
and then solve the resulting indicial equation (using the average of the two zeros 
for $z_d$) we get the exponent estimates
$0.833329270$ and $1.667679940$, which immediately suggests
that the leading exponent is $\theta =5/6$  in agreement with Dhar's result
and that possibly the sub-dominant exponent $\theta'$ is twice this.
The zero at $\simeq-0.1200$  though very close to $z_d$ could be distinct from $z_d$.
If we solve the indicial equation assuming an order 3 root we obtain the exponents
$0.000396421$, $0.836936143$ and  $1.688074312$, respectively, which clearly 
is no improvement on the order two assumption. Since the `new' exponent
is close to 0 this indicates that the assumption of a third order singularity
isn't well supported as this exponent could arise from an term analytic at $z_d$. 
We would expect a third actual exponent to by sub-dominant and
hence larger than $2\theta$. We note that if we look at fourth order approximants there 
is some evidence for a possible third order singularity that is three closely spaced
roots near $z_d$ with a third exponent $\theta''> \theta'$ but the evidence is not
very compelling.

\begin{figure}[htbp] 
   \centering
   \includegraphics[width=15cm]{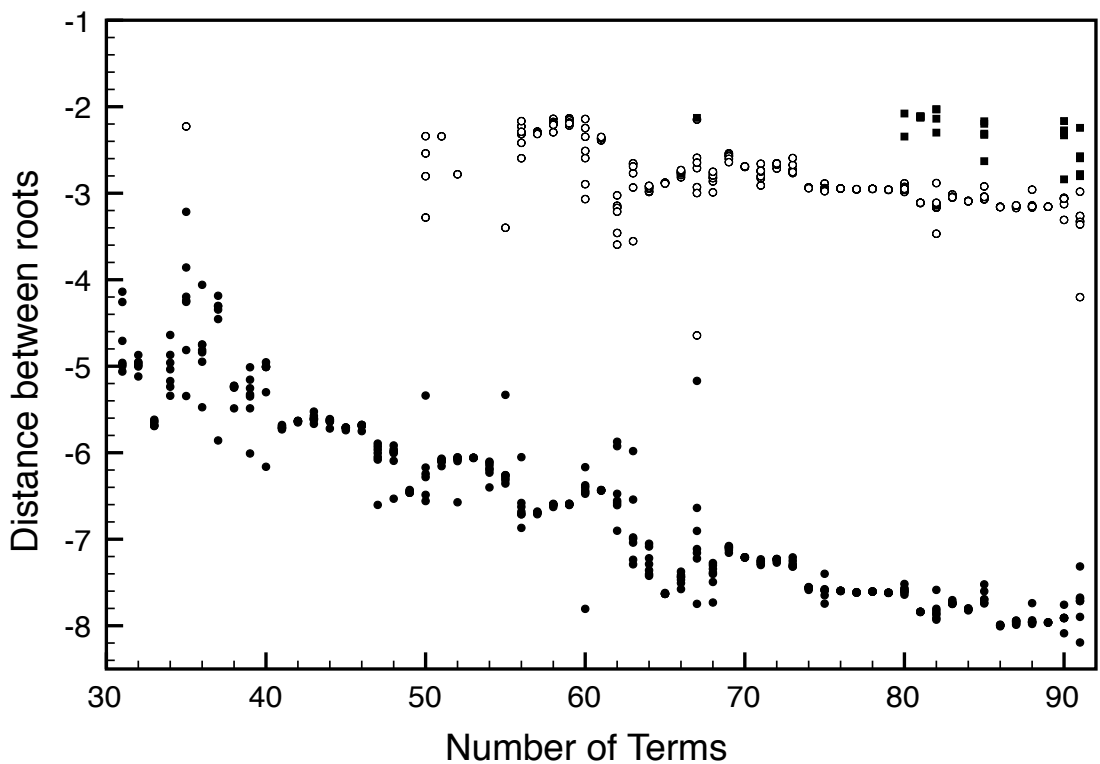} 
   \caption{ The distance between roots close to $z_d$ vs. the number of terms used to form 
   the homogeneous  third order differential approximant.}
   \label{fig:Roots}
\end{figure}

To further check the possible scenarios we calculated 
very many third order approximants by varying the degrees of the polynomials
appearing in (\ref{eq:diffapp}) and in each case we calculated all the roots of $Q_3(z)$ 
and then looked at any roots close to $z_d$. In figure~\ref{fig:Roots} we plot
the distance between roots against the number of terms used to form the approximants.
We have actually taken $log_{10}$ of the distance so the $y$-axis  indicates up to a sign
the number of digits the roots have in common. The solid circles shows  the distance between 
the closest lying roots, the open circles the distance to the next closest root, i.e. the zero at $\simeq-0.1200$ 
in table~\ref{tab:da21}, and finally the closed squares shows the distance to a possible fourth root. 
The possible third and fourth roots are only included if within 0.01 of $z_d$. 
From this plot it is clear that the series has at least a double root at $z_d$  since
all the approximants had at least two closely spaced roots and
the distance between these closest lying roots decreases monotonically as we increase
the number of terms  in the approximants. Further evidence for a double root is provided by
looking at  $Q_2(z)$. It the series truly has a double roots at $z_d$ then it follows that $z_d$
must appear as a root in $Q_2(z)$. We checked several of the approximants and in all cases
found that $Q_2(z)$ did indeed have a root very close to $z_d$. Obviously when using more than 60 or so terms
it seems that most approximants locate a third root close to $x_d$ and when the number
of terms go above 80 a fourth root appears. The distance from these roots to the
close pair appears to be weakly decreasing. This may indicate that the singularity at $z_d$
is actually of order grater than two.  However, given the distance from $z_d$ we can't
really confirm this numerically with any great confidence. 
  
 \begin{table}
   \caption{\label{tab:da21}   Real zeroes of $Q_3$ and
   the associated exponents obtained from a homogeneous third order differential approximant
   with polynomials of degree 21. }
   \begin{center}
 \begin{tabular}{cccc}
     \hline \hline
         Zero & Exponent  &  Zero & Exponent  \\ \hline
     3.797609243403    &   2.039756900951  & -0.302260961239    &   3.431304831739  \\
     3.696982277961    &  7.327149084651  & -0.379900722124    &   4.228776094690  \\
       {\bf -0.119338882393}    &   0.748746994128 &  -0.741300568486   &    2.959881511581 \\ 
     {\bf -0.119338893414}    &   2.752278507267   & -0.985745094861   &    1.581123317769 \\
     -0.120041211934    &   2.016623831595   &  -1.980070905154   &    2.502245197257 \\
     -0.173738685995    &   8.477340013372  &  -3.832447177296  &     0.541758005446  \\
    -0.259388829117    &   1.066025488201  \\
        \hline \hline
\end{tabular}
\end{center}
\end{table}

In table~\ref{tab:crpexp} we list estimates for $z_d$ and the two associated critical
exponents obtained by averaging several second or third order inhomogeneous 
differential approximants with a given degree $L$ of the inhomogeneous polynomial. 
The quoted error is simply the mean deviation of the approximants. 
We conclude that 
 
 $$z_d=-0.119338886(5), \quad \theta=0.83333(2), \quad \theta'= 1.6676(3).$$
Clearly $\theta =\frac56$ confirming Dhar's result.  $\theta'$ is tantalisingly 
close to   $2\theta$ and this is the most likely scenario.  While our current best estimate 
does seem to exclude this possibility  the numerical analysis is quite subtle and 
the `error-bars' on $\theta'$ should be viewed with some scepticism particularly since
the actual structure of the singularity at $z_d$ may be more complicated than assumed in our analysis.
As noted by Dhar  it seems that though the models of hard-squares and hard-hexagons
belong to different universality classes (the behaviour at $z_c$ is in the Potts universality
class with $q=2$ (Ising) and $q=3$, respectively)   the exponents at  $z_d$, appears to
be independent of the Potts index $q$ and for  hard-hexagond   Dhar has shown
that $\theta' =2\theta$. Indeed starting from Baxter's  \cite{Baxter80b,Baxter82} exact solution 
of the hard-hexagon model  Joyce \cite{Joyce88} showed that the low-density partition
function is a solution of an algebraic equation. From this one can in turn show that 
the partition function is the solution to a 12'th order ODE and at $z_d$ the  (non-analytical) 
exponents are $\frac56,\, \frac53,\,  \frac52,\ \frac{10}{3},\, \frac{25}{6},\, \frac{20}{3},\, \frac{15}{2},$ 
and $\frac{25}{3}$ \cite{Jensen12} , which clearly suggests that the structure of the singularity
 at $z_d$ for hard-squares is likely to be very complicated.

 \begin{table}
   \caption{\label{tab:crpexp}   Estimates for the dominant singular point
   $z_d$ and the associated exponents as obtained from second and third order
   differential approximants. $L$ is the degree of the inhomogeneous polynomial. The entry
   for $L=0$ are estimates from homogeneous approximants.}
   \begin{center}
 \begin{tabular}{llll}
     \hline \hline
     \multicolumn{4}{c}{Second order approximants} \\
      \hline \hline
         $L$ & \multicolumn{1}{c}{$z_d$} & \multicolumn{1}{c}{$\theta$} & \multicolumn{1}{c}{$\theta'$} \\ \hline
     0 & $-0.119338899(11)$  & $0.8333312(12)$ & $1.66864(62)$ \\
     2 & $-0.11933888907(24)$  & $0.83334442(22)$ & $1.667787(20)$ \\
     4 & $-0.1193388866(21)$ & $0.8333408(38)$ & $1.66760(16)$\\
     6 & $-0.1193388864(25)$ & $0.8333403(81)$ & $1.66753(23)$  \\
     8 & $-0.1193388868(18)$  & $0.8333425(35)$ & $1.66758(15)$ \\
  10  & $-0.11933888754(23)$ & $0.83334442(22)$ & $1.667643(20)$ \\ \hline
     \multicolumn{4}{c}{Third order approximants} \\
      \hline \hline
         $L$ & \multicolumn{1}{c}{$z_d$} & \multicolumn{1}{c}{$\theta$} & \multicolumn{1}{c}{$\theta'$} \\ \hline
    0 &  $-0.1193388890(18)$ & $0.8333289(10)$ & $1.66777(13)$ \\
     2 & $-0.1193388859(24)$ & $0.83320(30)$ &  $1.66745(25)$ \\
     4 & $-0.1193388866(24)$&  $0.83324(21)$ & $1.66753(28)$ \\
     6 & $-0.1193388849(33)$ & $0.83318(24)$ & $1.66735(38)$ \\
     8 & $-0.1193388847(36)$ & $0.83318(28)$ &$1.66733(40)$ \\
   10 & $-0.1193388837(16)$ & $0.83348(14)$ & $1.66718(19)$ \\
        \hline \hline
\end{tabular}
\end{center}
\end{table}

In conclusion we have analysed the series for $\kappa(z)$ using differential approximants
and found that the dominant singularity at $z_d$ appears to have order 2. When this
is taken into account the method of differential approximants  is perfectly well capable of
yielding accurate exponent estimates. In particular we confirm that $\theta=\frac56$ 
as found by Dhar \cite{Dhar83}.

 \section*{Acknowledgements}
 
The author was supported by the Australian Research Council via 
the Discovery Project grant DP120101593.

\end{document}